\documentclass[manuscript,screen]{acmart}

\usepackage{booktabs}
\usepackage{multirow}
\usepackage{xcolor}
\usepackage[table]{xcolor}
\usepackage{amsmath}
\usepackage[most]{tcolorbox}
\usepackage{array}
\usepackage{color}
\usepackage{subcaption}
\usepackage{booktabs}
\usepackage{xcolor}
\usepackage{colortbl}
\usepackage{amsmath}
\usepackage{siunitx}
\usepackage{multirow}
\usepackage{siunitx}
\usepackage{algorithm}

\usepackage{algorithm}
\usepackage{algorithmicx}
\usepackage{algpseudocode}
\usepackage{amsmath}

\usepackage{tcolorbox}   
\tcbset{
  colback=white,        
  colframe=black,       
  arc=2mm,              
  boxrule=0.5pt,        
  left=2mm, right=2mm,  
  top=1mm, bottom=1mm
}

\definecolor{lightpurple}{RGB}{240,240,255}
\definecolor{graytext}{gray}{0.5}
\setcopyright{none}
\renewcommand\footnotetextcopyrightpermission[1]{} 
\AtBeginDocument{%
  }

\begin{document}

\title{Completion by Comprehension: Guiding Code Generation with Multi-Granularity Understanding}


\author{Xinkui Zhao}
\authornote{Both authors contributed equally to this research.}
\email{zhaoxinkui@zju.edu.cn}
\affiliation{%
  \institution{Zhejiang University}
  \city{Hangzhou}
  \country{China}
}

\author{Rongkai Liu}
\authornotemark[1]
\email{15274800780@163.com}
\affiliation{%
  \institution{Zhejiang University}
  \city{Hangzhou}
  \country{China}
}
\author{Yifan Zhang}
\authornote{Corresponding author.}
\email{12451018@zju.edu.cn}
\affiliation{%
  \institution{Zhejiang University}
  \city{Hangzhou}
  \country{China}
}

\author{Chen Zhi}
\email{zjuzhichen@zju.edu.cn}
\affiliation{%
  \institution{Zhejiang University}
  \city{Hangzhou}
  \country{China}
}

\author{Lufei Zhang}
\email{zhanglf04@126.com}
\affiliation{%
  \institution{State Key Laboratory of Mathematical Engineering and Advanced Computing}
  \city{Zhengzhou}
  \country{China}
}

\author{Guanjie Cheng}
\email{chengguanjie@zju.edu.cn}
\affiliation{%
  \institution{Zhejiang University}
  \city{Hangzhou}
  \country{China}
}

\author{Yueshen Xu}
\email{ysxu@xidian.edu.cn}
\affiliation{%
  \institution{Xidian University}
  \city{Xian}
  \country{China}
}

\author{Shuiguang Deng}
\email{dengsg@zju.edu.cn}
\affiliation{%
  \institution{Zhejiang University}
  \city{Hangzhou}
  \country{China}
}

\author{Jianwei Yin}
\email{zjuyjw@cs.zju.edu.cn}
\affiliation{%
  \institution{Zhejiang University}
  \city{Hangzhou}
  \country{China}
}

\renewcommand{\shortauthors}{Xinkui Zhao et al.}

\begin{abstract}
As code completion task from function-level to repository-level, leveraging contextual information from large-scale codebases becomes a core challenge.
However, existing retrieval-augmented generation~(RAG) methods typically treat code as plain natural language, relying primarily on shallow semantic matching while overlooking structural semantics and code-specific dependencies. 
This limits their ability to capture control flow and underlying intent, ultimately constraining the quality of generated code.
Therefore, we propose \textbf{CoCo}, a novel framework that enables code \textbf{Co}mpletion by \textbf{Co}mprehension of multi-granularity context from large-scale code repositories.
CoCo employs static code analysis to extract structured context at the function, file, and project levels, capturing execution logic and semantic dependencies.
It then adopts an graph-based multi-granularity context selection mechanism to filter out redundant information and remove noise. 
Consequently, the information is converted into natural language in a consistent manner, thereby functioning as explicit contextual prompts to guide subsequent code completion.
Additionally, a structure-aware code re-ranker mechanism ensures alignment at both semantic and structural levels.
Extensive experiments on CrossCodeEval and RepoEval benchmarks demonstrate that CoCo consistently surpasses state-of-the-art baselines, achieving up to 20.2\% gains in EM.
Moreover, the framework is model-agnostic and can be seamlessly integrated into existing methods, leading to significant performance.
\end{abstract}





\maketitle

\section{Introduction}
Automatic code completion plays a fundamental role in modern software development~\cite{raychev2014code,robbes2008program,hellendoorn2019code,mcconnell2004code}, significantly boosting developer productivity and code quality~\cite{ziegler2022productivity,weber2024significant,muaruașoiu2015empirical}. 
Traditionally, code completion approaches have focused on generating code by leveraging static analysis~\cite{bruch2009learning,svyatkovskiy2019pythia}, rule-based heuristics~\cite{ye2002supporting}, or statistical language modeling~\cite{tu2014localness,hellendoorn2018deep,allamanis2018survey}.
While these methods perform well for simple scenarios—such as completing statements or standalone functions—they fall short in real-world settings where code is organized in complex repositories with abundant cross-file dependencies, customized APIs, and project-specific conventions~\cite{hellendoorn2017deep,raychev2016probabilistic}. 
The advent of large-scale pre-trained language models for code (Code LLMs), such as  GPT-4~\cite{achiam2023gpt}, DeepSeekCoder~\cite{guo2024deepseek,zhu2024deepseek}, Qwen~\cite{qwen2,hui2024qwen2} and Yi~\cite{young2024yi}, has greatly advanced the capabilities of automated code generation. 
However, these models are inherently limited by their input context window and cannot directly access the full scope of a repository, where critical information is often distributed across multiple files~\cite{wang2024rlcoder}.
This limitation becomes particularly challenging in \emph{repository-level code completion}, where accurate code generation requires a holistic understanding of repository-wide dependencies, shared utilities, and inter-module interactions. 
To address this challenge, the \emph{retrieval-augmented generation} (RAG) paradigm has emerged as a promising and widely adopted approach for the repository-level code completion task~\cite{yu2022bashexplainer,lu2022reacc}. 
Typically, RAG methods operate by first retrieving potentially relevant code snippets from a large codebase, and then concatenating these retrieved fragments with the original prompt before feeding the combined context into a code generation model. 
This mechanism enables the model to leverage external knowledge beyond the immediate file, thereby mitigating the limitations of input length and local context.


While RAG-based approaches enhance code generation by retrieving semantically similar code examples, they still face notable limitations.
First, most existing retrievers—whether lexical-based or model-based—treat code purely as text sequences and compute similarity based on surface-level token overlap, ignoring the inherent structural and semantic characteristics of source code.
As a result, retrieved examples often exhibit superficial lexical overlap, such as shared variable names or common keywords, with the query, yet fail to preserve deeper structural correspondences.
This structural mismatch introduces irrelevant or even misleading code fragments, which may divert the generation model from the intended logic, resulting in syntactically correct but semantically incorrect completions.
Second, retrieval results rarely capture fine-grained contextual dependencies within the same file or project-wide information scattered across multiple modules.
However, such information are often crucial for accurately completing complex code segments. 
The absence of this information during generation inevitably limits the model’s performance in real-world scenarios.

To address these problems, we propose \textbf{CoCo}, a novel framework that enables large language models to perform repository-level code completion by \textbf{comprehension before completion}.
Inspired by the way human developers first understand and analyze code context before writing new code, CoCo constructs a multi-granularity representation of the unfinished code to guide LLM-based generation.
Specifically, CoCo first applies static code analysis tools (such as AST parsers) to extract rich contextual information at the \emph{function}, \emph{file}, and \emph{repository} levels. 
At the function level, we analyze execution logic and control flow to capture the local semantics surrounding the unfinished code. 
At the file level, we identify both explicit and potential dependencies between the unfinished code and other entities in the same file. 
At the repository level, we model cross-file interactions and external module relationships.
Through these operations, we obtain a large volume of information related to the unfinished code. 
However, such information may be excessive, with some being unhelpful or even detrimental to guiding subsequent LLM-based code generation—there may also be redundancy or conflicts. 
To address this, CoCo incorporates an graph-based multi-granularity context selection mechanism to sift through and retain only the most relevant information.
The filtered information is then transformed into concise and coherent natural language descriptions, making the underlying structural and semantic context accessible to LLMs. 
Furthermore, CoCo introduces a structure-aware code re-ranker mechanism that refines retrieved code examples, ensuring both semantic relevance and structural consistency with the completion target.
Finally, CoCo synthesizes the multi-granularity contextual information and the retrieved code examples into a unified prompt, which is fed into the LLM for code completion.
Unlike conventional RAG-based methods that focus only on local code fragments, CoCo follows a ``\textbf{Comprehension first, then Completion}'' paradigm. 
This enables the LLM to reason globally and locally about cross-file dependencies, shared utilities, and project-specific conventions, resulting in more accurate and contextually consistent repository-level code completion.
Furthermore, our framework can be flexibly integrated with existing methods, further improving their performance by supplying high-level, structured context. 

Our main contrubution are:
\begin{itemize}
    \item \textbf{We propose CoCo, a novel comprehension-driven code completion framework.} CoCo enables large language models to perform repository-level code completion by deeply comprehending the context of unfinished code, rather than relying solely on fragment-level retrieval.

    \item \textbf{We design a multi-granularity analysis pipeline} that gathers and filters function-, file-, and repository-level contextual information and transforms the resulting insights into LLM-friendly natural-language prompts.

    \item \textbf{We introduce a structure-aware code re-ranker mechanism} that considers both semantic similarity and structural consistency, ensuring that retrieved code exemplars are highly relevant and contextually aligned with the unfinished code.

    \item \textbf{We demonstrate the effectiveness and generalizability of CoCo through extensive experiments on both CrossCodeEval and RepoEval.} CoCo consistently outperforms state-of-the-art methods and integrates flexibly with existing frameworks. Our code is publicly available. We make the code publicly available at \href{https://anonymous.4open.science/r/code-generation-6B63}{the link}.

\end{itemize}

\section{Background}

\subsection{Retrieval-Augmented Generation Enhanced Repository Code Completion}

RAG~\cite{gao2023retrieval,lewis2020retrieval} is a general framework that enhances LLMs by leveraging external knowledge retrieved from a large corpus. 
A typical RAG workflow consists of three main stages: (1) \textbf{Indexing}: The external corpus is first pre-processed and encoded into a searchable index, often using dense or sparse vector representations; (2) \textbf{Retrieval}: Given an input query or prompt, the retriever searches the index to select relevant documents or knowledge pieces; (3) \textbf{Generation}: The retrieved information is then incorporated into the model's context and the LLM generates the final output based on this augmented context. By integrating relevant external knowledge, RAG enables LLMs to overcome the limitations of their parametric memory and adapt more flexibly to downstream tasks.

In recent years, researchers have conducted a substantial
amount of research  utilizing RAG for code-related research \cite{lu2022reacc,bui2024rambo, mansur2024ragfix, liu2024graphcoder, okutan2024leveraging, yu2022bashexplainer, li2021editsum, ding2025retrieval, zhang2024raft, mukherjee2025sosecure}. 
For repository-level code generation, directly providing the entire codebase as context to LLMs is infeasible due to the massive size of repositories and the limited context window of LLMs~\cite{wang2024rlcoder}. Moreover, code fragments to be generated often share similarities or semantic connections with existing code within the repository. 
And traditional LLMs for code generation are often constrained by the closed world of their training data, making it difficult to access up-to-date API usages, third-party library information, or relevant private knowledge within code repositories. As a result, RAG has proven to be an effective solution in this domain. Typically, RAG methods selectively retrieve relevant code snippets and inject them into the model's context window. This targeted retrieval enables the LLM to access both intra-file and cross-file information~\cite{liao2023context,ding2022cocomic}, thereby enhancing its ability to generate more accurate and contextually appropriate code completions.

\subsection{Limitations}

Despite its promise, this approach is subject to several inherent limitations. Conventional RAG methods typically assume that, for any code completion task, sufficiently similar and relevant code snippets can always be found within the repository. In practice, however, this assumption frequently fails, giving rise to several fundamental challenges.

\begin{figure}[h]
    \centering
    \includegraphics[width=1\linewidth]{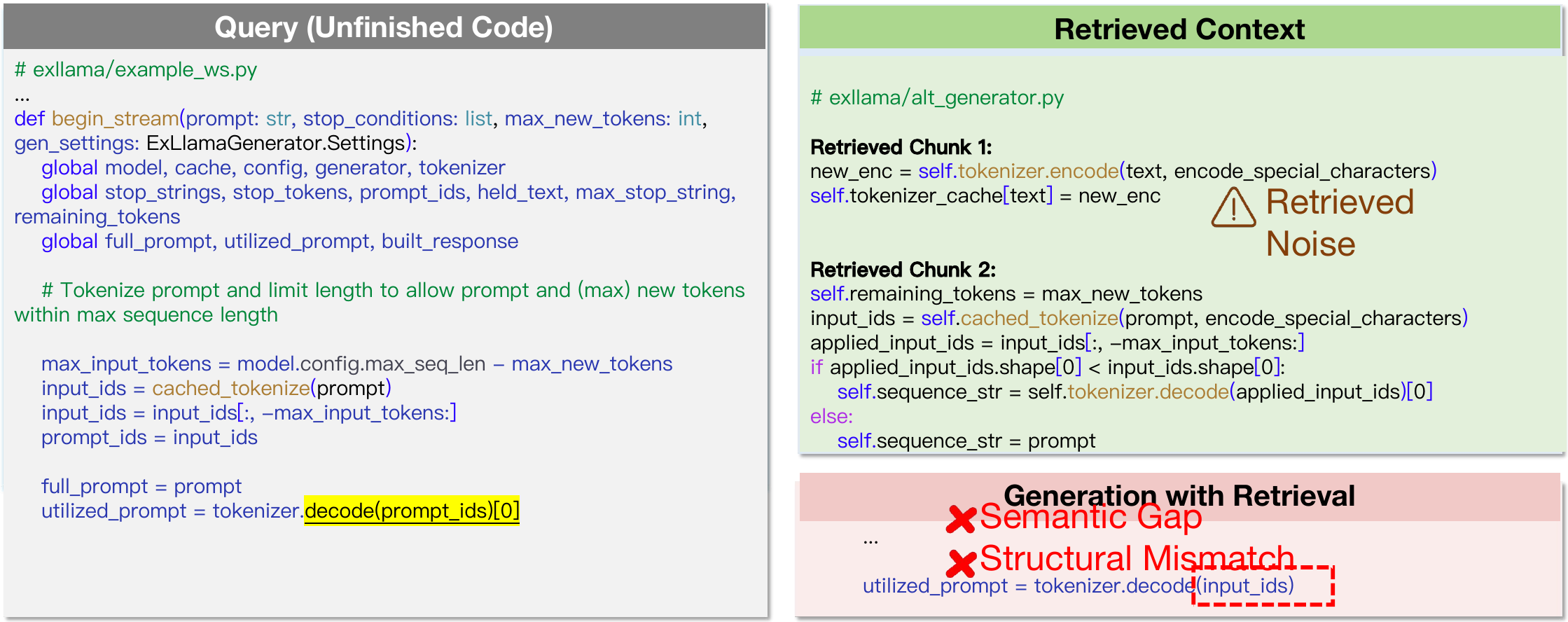}
    \caption{Limitations of traditional retrieval-based methods can introduce semantic and structural context gaps, leading to the generation of incorrect code.}
    \label{fig:limitations}
\end{figure}

\textit{First}, traditional retrieval methods, such as lexical-based approaches BM25~\cite{robertson2009probabilistic,jimenez2023swe}, generally treat code as plain text and struggle to capture the underlying semantics and intent of code snippets. Such a semantic gap can cause the retriever to select code that appears textually similar but is actually semantically irrelevant, resulting in ineffective or even misleading augmentation for code generation. For example, as illustrated in Fig.~\ref{fig:limitations}, the retriever selects a lexically similar variable \texttt{input\_id} by mimicking \texttt{applied\_input\_ids} without considering the actual semantic meaning of these variables.
\textit{Second}, retrieval noise and errors are difficult to eliminate entirely: the retriever may include irrelevant, redundant, or misleading code snippets, and incorporating such noisy context into the prompt can confuse the generation model and degrade completion quality. 
\textit{Third}, most existing retrieval mechanisms inadequately model the complex structural and cross-file dependencies present in real-world repositories, such as data flows, control flows, and inter-file relationships. This structural context gap makes it challenging to provide the generation model with the most relevant and informative context for accurate code completion. Another factor contributing to the failure in Fig.~\ref{fig:limitations} is the neglect of the grammatical and structural roles of variables and functions in the context. Variables like \texttt{input\_ids} often serve as consumed intermediate variables: they are used immediately after their definition and are seldom directly referenced again in the remaining code. Due to the lack of understanding of code control flow and syntactic structure, the retriever or generation model tends to select such already-consumed variables for completion, rather than prioritizing other unused-but-defined variables that are more likely to be semantically appropriate in the current context.

\section{Methodology}
\subsection{Overview}

\begin{figure}[ht]
    \centering
    \includegraphics[ height=0.3\textheight]{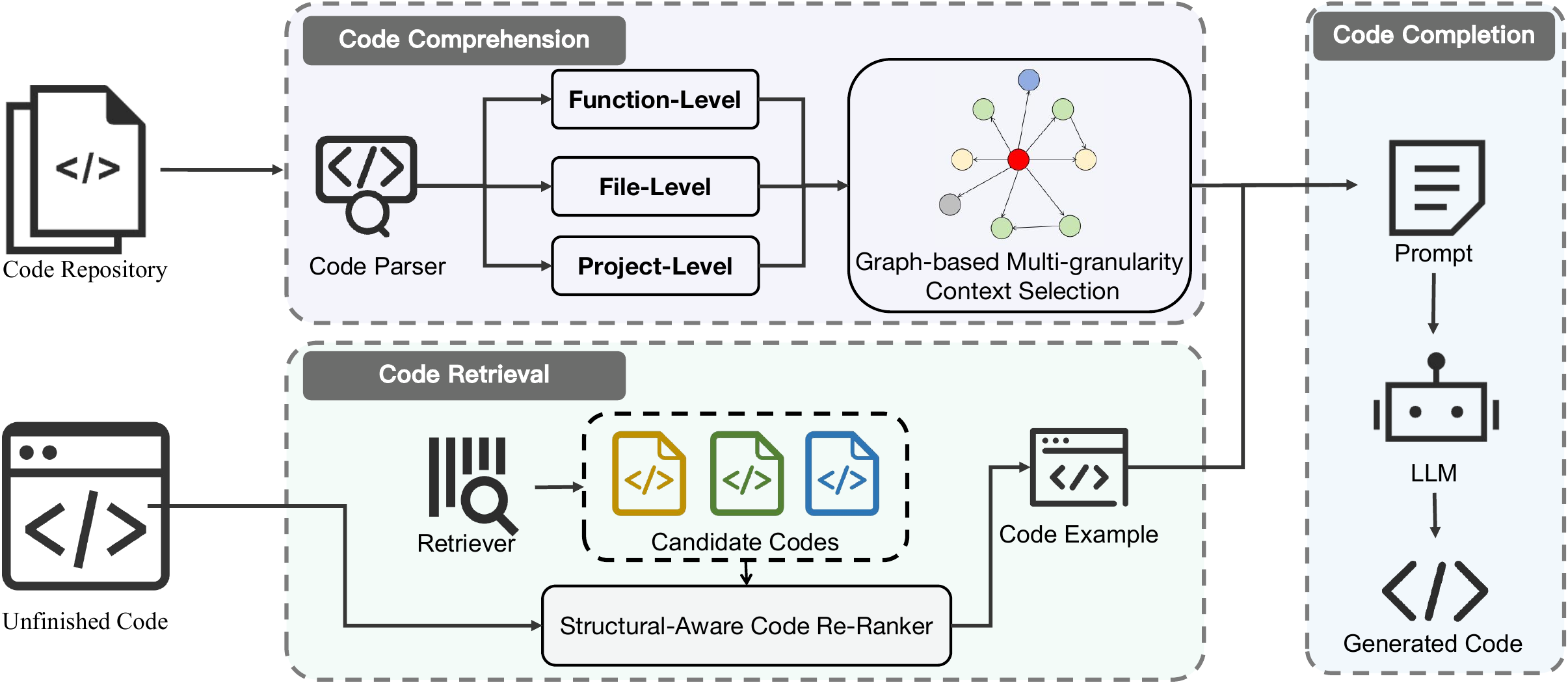}
    \caption{Oveiview of CoCo.}
    \label{fig:overview}
\end{figure}

In this section, we introduce CoCo, a framework for repository-level code completion that captures multi-granularity context to guide LLMs in generating code that is aligned with both semantic intent and structural constraints.
As illustrated in Fig.~\ref{fig:overview}, CoCo is composed of three modules: code comprehension, code retrieval, and code generation.
In the code comprehension stage, we use static analysis tools to extract background information relevant to the unfinished code, including execution logic, in-file dependencies, and cross-file module references. 
To ensure the resulting information is not overly redundant and does not negatively impact subsequent generation tasks, we employ a graph-based multi-granularity context selection mechanism to sift through the data, retaining only the most critical and relevant information.
This information is then unified and structured into a context representation that improves the LLM's comprehension of the intended logic flow and developmental intent.
In the code retrieval stage, we retrieve similar code snippets from the code repository that are semantically similar to the unfinished code.
To ensure that the retrieved candidates are not only semantically relevant but also structurally consistent, we further apply a structure-aware code re-rank mechanism to refine the results.
Finally, in the code generation stage, the contextual information collected is integrated into a single prompt. 
This prompt is then fed into the LLM to generate the continuation code that complies with the given context.
\subsection{Code comprehension}

In the task of repository-level code completion, the code repository itself contains rich structural and semantic information that can inform the generation of unfinished code. 
To fully exploit such information, we employ static analysis tools to construct multi-granularity contextual representations surrounding the unfinished code, enabling the model to reason about potential control flows and developer intent, rather than relying solely on surface-level similar code examples.
Specifically, we utilize Tree-sitter and the abstract syntax tree (AST) module to dynamically zoom out from the function level, progressively expanding the context to encompass the file and project levels. This process forms a multi-granularity representation grounded in execution logic, semantic dependencies, and inter-module relations.


\begin{figure}[h]
    \centering
    \includegraphics[width=0.7\linewidth]{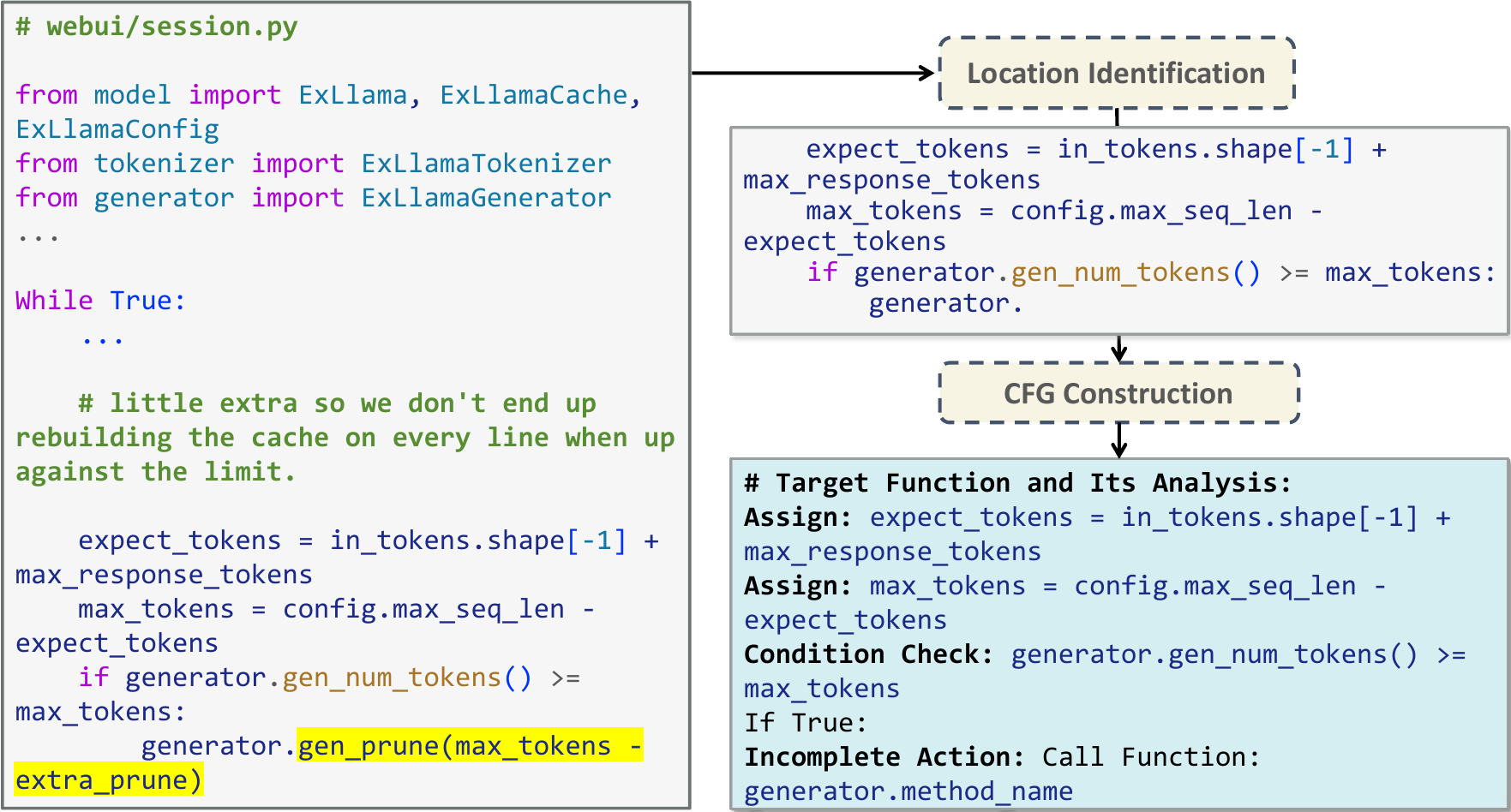}
    \caption{Illustration of Function-Level Analysis. The yellow code block indicates unfinished code.}
    \label{fig:func}
\end{figure}


\subsubsection{Function-Level Analysis.}
Within a function, the surrounding code typically provides the most relevant context for the missing line, containing the majority of information needed for accurate code completion. 
This is because local variables, control flows, and intermediate computations within the same function directly influence the semantics of the target code. 
To effectively capture these local execution semantics, we introduce a function-level analysis mechanism that constructs a control-flow-aware context preceding the target line.
This enables the large language model to better understand the semantic structure and reason about the logical pattern of the unfinished code.
We begin by constructing the AST of the source file and recursively traversing its nodes to locate the target function of the target line. 
Specifically,  the \texttt{function\_definition} node is identified as the target function if its start and end line numbers encompass the line where the code is missing.
Once the target function is located, we extract the code segment ranging from the beginning of the function up to the target line as the local context for code generation.
If the target line does not reside within any function body—i.e., it appears at the top-level script scope—we instead extract the code before the target line as local context.
As illustrated in Fig.~\ref{fig:func}, after obtaining the local code context, we construct a control flow graph~(CFG) to model the control dependencies among statements. 
During this process, we identify key control structures within the AST, such as conditionals~(\texttt{if}), loops~(\texttt{for, while}), and termination statements~(\texttt{return}). 
These statements are treated as nodes in the graph. 
Directed edges are then added between nodes based on the program’s execution semantics to reflect control flow transitions.
This CFG-based structural representation makes the reachable paths prior to the target line explicit, thereby providing the language model with a clearer execution context to guide code generation.

\begin{figure}[h]
    \centering
    \includegraphics[width=0.7\linewidth]{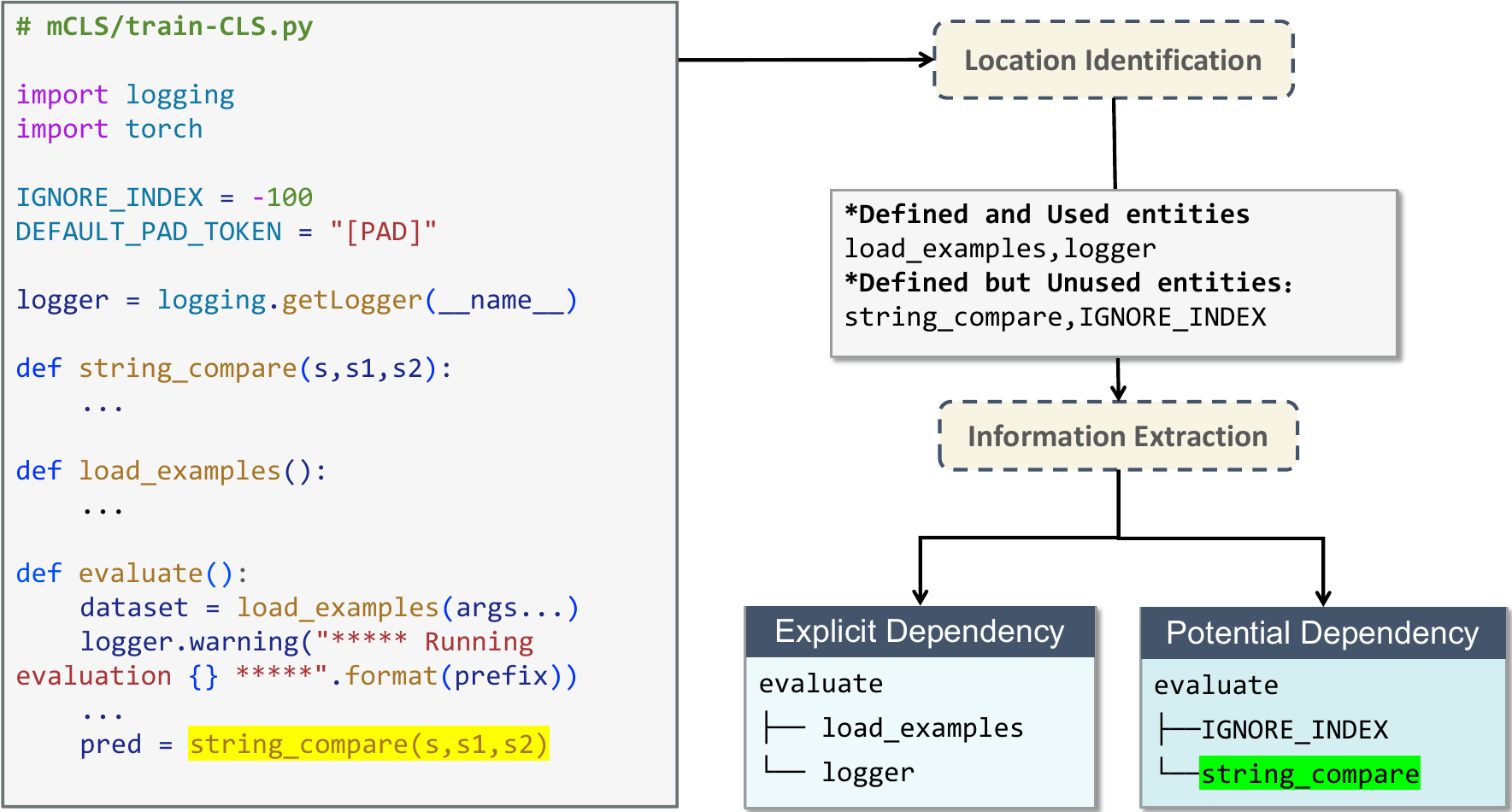}
    \caption{Illustration of File-Level Analysis. The yellow code block indicates unfinished code. The green code block indicates useful information.}
    \label{fig:file}
\end{figure}
\subsubsection{File-Level Analysis.}
To enhance the semantic coverage of the context, we also introduce a file-level analysis mechanism to model the dependencies between the target code and other entities defined within the same source file. 
Specifically, we first parse the abstract syntax tree of the source file and traverse upward from the target line to collect all functions and variables defined before the target line, thereby constructing a static symbol definition set.
We then extract all identifiers used in the target function to form a symbol usage set.
By computing the intersection between the definition and usage sets, we extract the explicit dependency information~(while filtering out function-internal self-references such as local variables and recursive calls), which precisely captures the target function's actual reliance on other entities within the file. 
This information includes the dependent symbols' names, types, locations, and corresponding source code snippets. 
When the unfinished code is located outside any function body (i.e., in a script-level context), we skip dependency matching and extract only symbols.
In addition to explicit dependencies, we also identify potential dependency information—symbols that are defined in the current file but have not yet been used prior to the target line. 
This is accomplished by computing the set difference between the sets of defined and used symbols, a process applied regardless of whether the missing code appears within a function or at the top level. 
Although these symbols have not been referenced so far, they are often intended for use in subsequent code, reflecting the programmer’s design intent. 
As illustrated by the green-highlighted block in the lower-right corner of Fig.~\ref{fig:file}, \texttt{string\_compare} remains unused in the existing code but is invoked in the incomplete statement, demonstrating the importance of potential dependency information in facilitating accurate code completion.

\begin{figure}[h]
    \centering
    \includegraphics[width=0.7\linewidth]{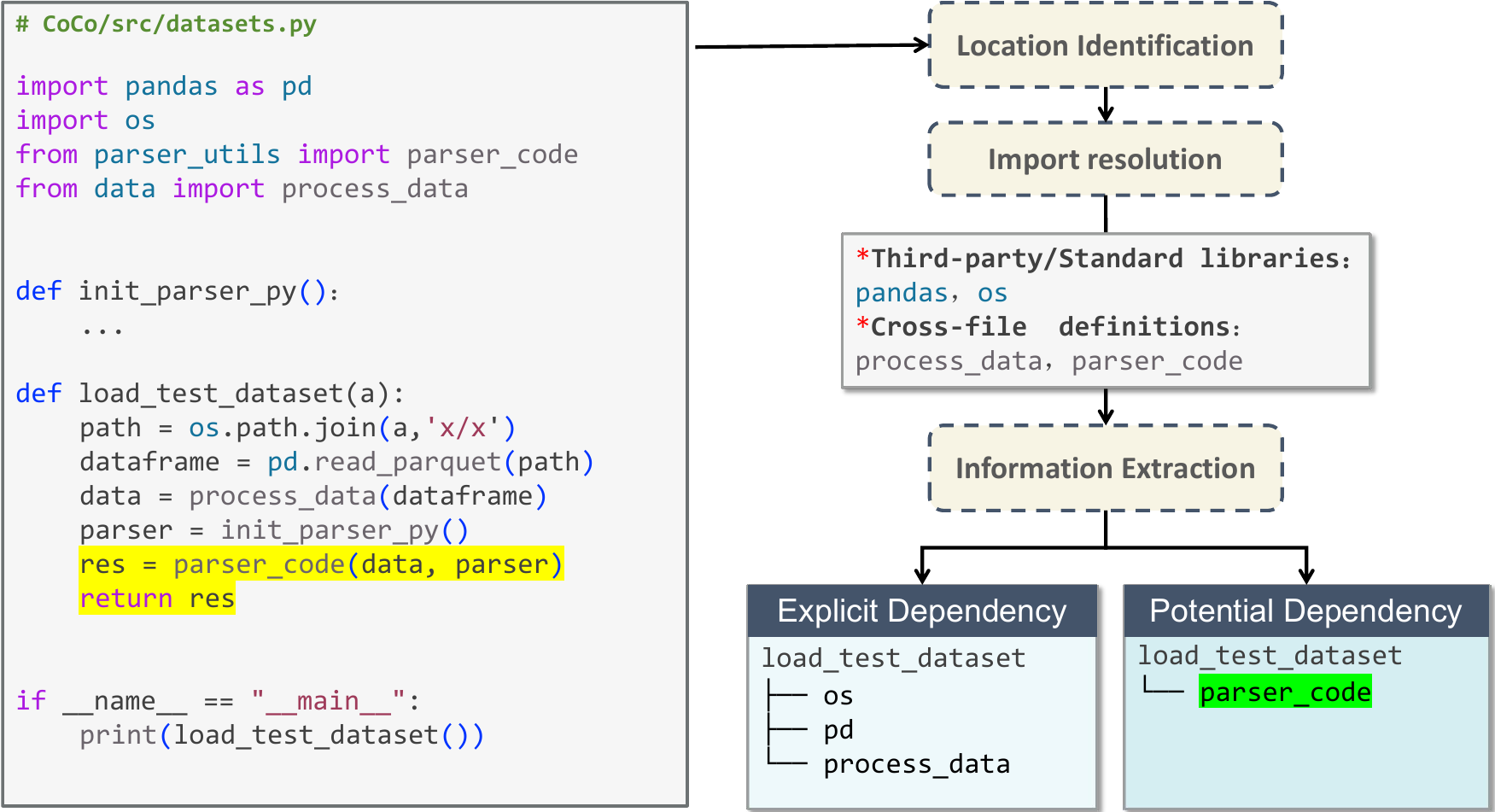}
    \caption{Illustration of Project-Level Analysis. The yellow code block indicates unfinished code. The green code block indicates useful information.}
    \label{fig:pro}
\end{figure}

\subsubsection{Project-Level Analysis.}
Real-world code repositories frequently exhibit complex interactions across multiple modules and external libraries. Accurately modeling these cross-file and cross-library dependencies is critical for providing comprehensive context and enabling the generation model to produce semantically correct and contextually appropriate code completions.
To capture dependencies between the target code and external modules or libraries, we introduce a project-level code analysis mechanism that enhances contextual completeness. 
This mechanism focuses on two categories of critical information. The first is explicit cross-module dependencies, which refer to entities that are both imported and actively referenced within the target function—either from other source files in the project or from external libraries. The second is potential cross-module dependencies, which denote symbols that have been imported into the current file but have not yet been used before the target line; these may be invoked in subsequent code and thus are highly relevant for code completion.

We begin by parsing the AST of the target file to extract all import statements, categorizing them into cross-file definitions (i.e., symbols defined in other source files within the same repository) and third-party or standard library imports. To distinguish between these, we recursively scan all Python files in the project root to build a mapping from module names to their corresponding physical file locations. If an imported module matches an entry in this mapping (via full-name or suffix matching), it is classified as a cross-file definition; otherwise, it is treated as a third-party or standard library.
For cross-file definitions, we further parse the corresponding source files to extract the imported function or class definitions, recording their names, types, source code locations, and full implementations. 

Next, we extract the set of identifiers used within the target function and match them against the previously collected imported definitions. All matched entities constitute the set of explicit cross-module dependencies, which may include both cross-file symbols and external libraries. Dependencies are organized by their source: for cross-file definitions, we record their source locations and code content in a cross-file dependency dictionary; for third-party or standard libraries, only the module and symbol names are retained.
As illustrated in Fig.~\ref{fig:pro}, the collected dependencies include \texttt{os} (a standard library), \texttt{pd} (a third-party library) and \texttt{process\_data} (a cross-file definition), together forming the complete set of explicit cross-module dependencies. If the unfinished code is not located within a function body, identifier matching is omitted and only definitions are extracted.

Building on this, we further identify potential cross-module dependencies—imported symbols that have not yet been used before the target line but are likely to appear in subsequent code. Such symbols often reflect the programmer’s latent intent and are valuable for guiding the model in completing the unfinished code. Specifically, we compute the set difference between all imported entities and those symbols already used (regardless of whether they occur within a function body). The resulting items are treated as potential dependencies, for each of which we record the import statement’s source code location, dependency type, and, if applicable, the symbol’s definition location and code snippet.
As shown in the green highlighted area in the lower-right of Fig.~\ref{fig:pro}, although \texttt{parse\_code} is not used before the target line, it is invoked in the unfinished statement, demonstrating the importance of potential cross-module dependency information in facilitating accurate code completion.

\subsubsection{Graph-based Multi-granularity Context Selection}
Although multi-granularity analysis provides a comprehensive view of the unfinished code, directly feeding all extracted information into a large language model is impractical. 
File-level and project-level analyses often generate substantial amounts of data. Incorporating all such information inevitably exceeds the model’s token budget, while naïve truncation risks removing semantically crucial elements, thereby damaging contextual integrity and resulting in generated code that no longer aligns with the surrounding program structure. 
Meanwhile, the raw context frequently contains redundant or irrelevant elements.
These noisy components not only fail to support the generation process but may also mislead the model by diluting the semantic signals associated with the unfinished code.
To address these challenges, we introduce a graph-based context selection mechanism that systematically models the semantic relations among multi-granularity elements and quantifies their relative importance.
The detailed algorithmic procedure is shown in Algorithm~\ref{alg:distillation}.

We first construct a heterogeneous semantic graph from the parsed multi-granularity context. 
The target function containing the unfinished code is designated as the central node, while additional nodes represent symbols, type definitions, calls, import statements, and cross-file entities extracted from file- and project-level analyses.
Edges are constructed strictly according to semantic relations present in the source code, such as function invocations, variable usages, and inheritance relations, ensuring that the graph accurately reflects the program’s underlying structure.
We add directed edges from the central node to all other nodes so that the ranking process is explicitly centered on the function under completion and importance propagation naturally originates from it.

We compute importance scores using Personalized PageRank, which differs from classical PageRank by allowing the random walk to restart preferentially from a designated subset of nodes. Given a graph 
$G=(V,E)$, the importance score for node $v_i$ is defined as:

\begin{equation}
    Score(v_i) = \alpha   \sum_{v_j\in In(i)}^{}\frac{Score(v_j)}{deg(v_j)} + (1-\alpha ) p(v_i)
\end{equation}
where $\alpha$ is the teleportation factor, $In(i)$ denotes the set of nodes with edges pointing to $v_i$, $Score(v_j)$ is the PageRank score of node $v_j$ in the previous iteration, $deg(v_j)$ is its out-degree, and $p(\cdot)$ is the personalization distribution. 

In our setting, $p(v)$ assigns probability 1 to the central node and 0 to all other nodes, meaning that random-walk restarts always return to the target function rather than choosing uniformly at random. 
This ensures that importance diffuses outward from the missing-code function and decays with structural distance.
After convergence, we rank file-level and project-level nodes separately and select the top-k nodes from each level as the distilled context for code generation.

\begin{algorithm}[H]
    \centering

    \begin{algorithmic}[1]
        \Require
            File-level, Project-level parsed context and target function $f$
        \Ensure
            Distilled context (top-$k$ file/project-level nodes)
        
        \State Construct heterogeneous semantic graph $G=(V,E)$ from parsed context
        %
        
        
        \State Compute node importance scores using Personalized PageRank on $G$:
        \State \quad For all $v_i \in V$, iteratively update until convergence:
        \State \qquad $\text{Score}(v_i) = \alpha \sum_{v_j \in \text{In}(i)} \frac{\text{Score}(v_j)}{\deg(v_j)} + (1-\alpha) p(v_i)$
        \State \qquad $p(v) = \begin{cases} 
                            1 & \text{if } v = v_{\text{central}} \\ 
                            0 & \text{otherwise} 
                          \end{cases}$
        
        \State Split nodes into file-level set $V_{\text{file}}$ and project-level set $V_{\text{proj}}$
        \State Sort both sets by their importance scores in descending order
        \State Select top-$k$ nodes from each sorted set to form the distilled context
        
        \State \Return $\text{distilled context}$  
    \end{algorithmic}
    \caption{Graph-based Multi-granularity Context Selection and PPR}
    \label{alg:distillation}
\end{algorithm}

\subsection{Code Retrieval}
In the code retrieval stage, we follow the common RAG paradigm to retrieve semantically similar code snippets from the repository as examples to guide subsequent code generation. 
However, existing methods typically treat code as plain text, relying solely on embedding-based semantic similarity, which may result in the retrieval of structurally mismatched examples and thereby impair generation quality.
To address this limitation, we introduce a Structure-Aware Code Re-Ranker that enhances the structural consistency of retrieved exemplars. 
Specifically, after obtaining an initial set of candidate snippets via RAG, we extract AST path representations from both the query and each candidate, and compute a structure score based on the Jaccard similarity.
The calculation formula is as follows.
\begin{equation}
    structure\_score = \frac{\left | P_{query}\cap P_{candidate} \right | }{\left | P_{query}\cup  P_{candidate} \right |} 
\end{equation}
where $P_{query}$ and $P_{candidate}$ denote the sets of AST paths extracted from the query and candidate snippets, respectively.
The final ranking score is computed as a weighted combination of the semantic and structural scores. 
Based on the final score, we re-rank the candidate snippets and select the top\-k results as retrieval examples.
\begin{figure}[ht]
    \centering
    \includegraphics[height=0.3\textheight]{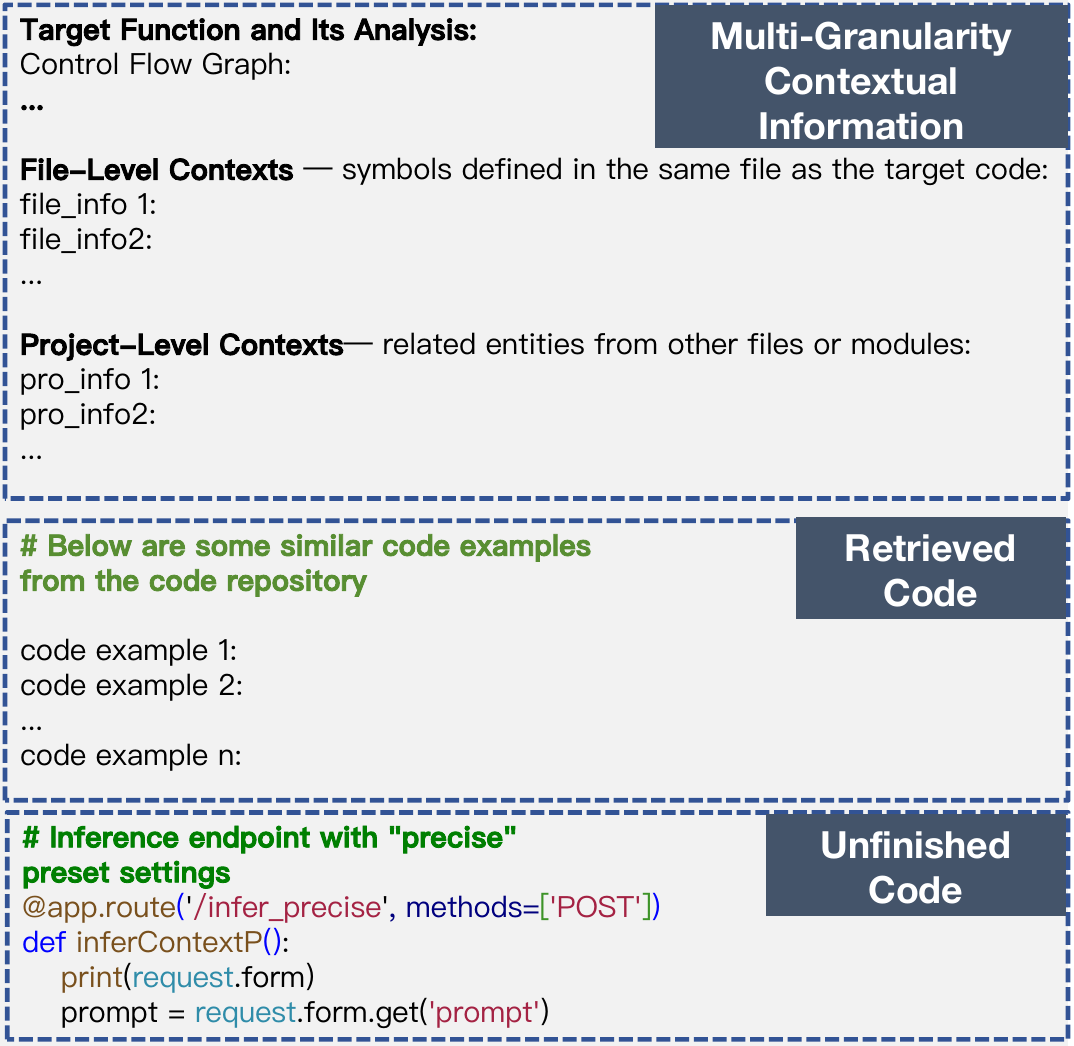}
    \caption{Prompt template provided to the code generation model, consisting of (1) multi-granularity contextual information that captures control flow, file-level, and project-level code relationships; (2) retrieved similar code examples from the repository; and (3) the unfinished code segment to be completed. * represents the optional context.}
    \label{fig:prompt}
\end{figure}

\subsection{Code Generation}
In this stage, we construct a unified prompt by integrating multi-granularity contextual information and similar code examples, which serves as the direct input to the LLM for code generation.
As shown in Fig.~\ref{fig:prompt}, the prompt enhances existing prompting strategies by incorporating function-level, file-level, and project-level contextual information.
It is structured into three components: multi-granularity contextual infromation, similar code examples, and the target code to be completed.

\section{Experimental Setup}
In this section, we will introduce the experimental environment, the evaluation metrics adopted, the compared methods, and the analysis of the experimental results.
And we aim to answer the following research questions through our experimentation.
\begin{itemize}
    \item \textbf{RQ1.} How effective is CoCo in repository-level code completion?
    \item \textbf{RQ2.} Does each component of CoCo contribute to its performance?
    \item \textbf{RQ3.} How well does CoCo generalize as a plug-in to other methods in terms of performance? 
    \item \textbf{RQ4.} What is the process latency of integrating CoCo, and is it acceptable relative to its performance gains?
    \item \textbf{RQ5.} How do different levels of contextual granularity affect code completion performance?
\end{itemize}

\subsection{Benchmark}
To verify the effectiveness of CoCo, we select two widely used benchmarks for comparative experiments: CrossCodeEval~\cite{ding2023crosscodeeval} and RepoEval~\cite{zhang2023repocoder}.
CrossCodeEval is a multilingual repository-level code generation benchmark, where the generation of each sample need rely on cross-file information. 
It covers four programming languages: Python, Java, TypeScript, and C$\#$. 
We use the Python and Java subsets for the experiments.
RepoEval is a multigranular repository-level code generation benchmark, containing subsets at three levels: line, api, and function, which covers scenarios of real-world repositories. 
We use the line-level and api-level subsets for the experiments, which consist of 1600 samples.

\subsection{Baselines}
To verify the effectiveness of CoCo, we select some commonly used and state-of-the-art~(sota) methods to compare, including RawRAG, RepoCoder, and RLcoder.

\begin{itemize}
    \item \textbf{RawRAG.} It is a standard repository-level code generation method, consisting of two parts: similar code retrieval and code generation. The process is as follows: retrieve code snippets similar to the code to be generated through the retriever, then concatenate these snippets with the code to be generated into a prompt, and input it into a LLM for code completion.
    \item \textbf{RepoCoder.} As a commonly used repository-level code generation method, it shares the same overall workflow as RawRAG. However, it has improved the code retrieval process by adopting an iterative retrieval approach to enhance the quality of similar code.
    \item \textbf{RLCoder.} It is a sota method in the field of repository-level code generation, which also focuses on optimizing the code retrieval process. It trains the retriever using a reinforcement learning mechanism and eliminates retrieval candidates that may have negative impacts through a stop signal mechanism.

\end{itemize}

\subsection{Metrics}
We use four widely adopted evaluation metrics to compare and analyze the experimental results: Exact Match~(EM), Edit Similarity~(ES), Identifier Exact Match~(ID.EM) and F1-score.
\begin{itemize}
    \item EM measures the proportion of generated code that exactly matches the ground-truth code, quantifying overall correctness, which calculation process is as follows.
    \begin{equation}
        EM = Judge(y_i,y^*_i)
    \end{equation}
    where $y$ and $y^*$ represent the generated code and ground-truth code respectively, $Judge()$ refers a function that determines whether two codes are equal.
    
    \item ES evaluates the similarity between generated and ground-truth code, assessing statement matching granularity, which calculation process is as follows.
    \begin{equation}
        ES = 1 - \frac{Lev(y_i, y^*_i)}{max(len(y_i), len(y^*_i))} 
    \end{equation}
    where $Lev()$ denotes the Levenshtein distance, which quantifies the edit distance between two codes, and $len()$ represents a function that calculates the string length.

    \item ID.EM measures the proportion of generated code identifiers that match the ground-truth code. Its calculation process is similar to EM, except that the object is replaced by the set of identifiers.

    \item F1 comprehensively evaluates the model's performance in predicting core identifiers in code through the harmonic mean of precision~(the proportion of correctly predicted identifiers among all predicted ones) and recall~(the proportion of truly existing identifiers that are correctly predicted).

\end{itemize}

\subsection{Experiment Settings}
All experiments are conducted on a machine with one Tesla A100 GPU, with 40 GB memory. 
In the experments, we select three different LLMs as code generation backbone, including DeepSeekCoder-1B~\cite{guo2024deepseek}, Codellama-7B~\cite{roziere2023code},Yi-Coder-1.5B~\cite{young2024yi}. The selection of these models is motivated by our goal to comprehensively evaluate the effectiveness of our framework across different model sizes and architectures. Notably, we place particular emphasis on the performance of smaller-scale models, as their efficiency and deployment-friendliness are of great practical significance for real-world applications and resource-constrained scenarios.
Following the setup of other methods, the random seed is set to 123 and the max number of generated tokens is limited to 64 for all models.
And we use the retriever of RLCoder~\cite{wang2024rlcoder} as our retriever.
To ensure fair and consistent evaluation, we directly reuse the publicly available implementations of evaluation metrics from previous studies~\cite{ding2023crosscodeeval,zhang2023repocoder}, rather than reimplementing them.

\begin{table*}[h]
\centering
\definecolor{myblue}{rgb}{0.9, 0.95, 1.0}
\caption{Performance comparison of different models and baselines. CL-7B denotes CodeLlama-7B, Yi-1.5B denotes Yi-Coder-1.5B, and DSC-1B denotes DeepSeekCoder-1B.}
\label{tab1}
\renewcommand{\arraystretch}{1.3}
\setlength{\tabcolsep}{2pt}
\scalebox{0.9}{
\begin{tabular}{lcccccccccccccccc}
\toprule
\multirow{2}{*}{\textbf{Method}}
& \multicolumn{4}{c}{\textbf{CrossCodeEval (Python)}} & \multicolumn{4}{c}{\textbf{CrossCodeEval (Java)}} & \multicolumn{4}{c}{\textbf{RepoEval (Line)}} & \multicolumn{4}{c}{\textbf{RepoEval (API)}} \\
\cmidrule(lr){2-5} \cmidrule(lr){6-9} \cmidrule(lr){10-13} \cmidrule(lr){14-17}
& \textbf{EM} & \textbf{ES} & \textbf{ID.EM} & \textbf{F1} & \textbf{EM} & \textbf{ES} & \textbf{ID.EM} & \textbf{F1} & \textbf{EM} & \textbf{ES} & \textbf{ID.EM} & \textbf{F1} & \textbf{EM} & \textbf{ES} & \textbf{ID.EM} & \textbf{F1} \\
\midrule

RawRAG {\footnotesize\textit{CL-7B}}  & 20.56 & 67.34 & 29.86 & 59.01  &20.10 &63.76 &27.58 &55.99 & 42.87 & 64.68 & 27.37 & 47.40 & 34.68 & 62.06 & 28.50 & 60.89  \\
RepoCoder {\footnotesize\textit{CL-7B}}   & 24.24 & 69.73 & 34.26  & 61.11  &22.21 &63.83 &30.48 &56.79 & 41.88 & 64.28  & 26.75  & 47.29  & 37.25 & 63.57  & 30.06  & 62.34  \\
RLCoder {\footnotesize\textit{CL-7B}} & 27.17 & 71.68 & 38.19  & 64.44  &23.79 &64.81 &32.58 &58.12 & 48.00 & 69.01 & 29.12 & 49.87  & 37.68 & 64.06 & 30.62  & 62.56  \\
\rowcolor{myblue}
\textbf{CoCo {\footnotesize\textbf{\textit{CL-7B}}}} & \textbf{31.23} & \textbf{74.87} & \textbf{41.95} & \textbf{67.89} &\textbf{26.21} &\textbf{66.47} &\textbf{35.17} &\textbf{62.29} & \textbf{49.93} & \textbf{69.88} & \textbf{30.71} & \textbf{50.75} & \textbf{41.21} & \textbf{67.37} & \textbf{33.14} & \textbf{65.93} \\

\midrule

RawRAG {\footnotesize\textit{Yi-1.5B}} & 18.20 & 66.72 & 27.39  & 55.93  &15.52 &63.14 &24.03 &54.49 & 38.75   & 61.26  & 24.94  & 45.02 & 32.50   & 60.10   & 27.06   & 60.19   \\
RepoCoder {\footnotesize\textit{Yi-1.5B}} & 21.73 & 69.01 & 31.63  & 59.64  & 16.92&63.98 &25.62 &55.58 & 41.50  & 63.35  & 26.38  & 46.58  & 35.00  & 62.15  & 28.25 & 61.43  \\
RLCoder {\footnotesize\textit{Yi-1.5B}}& 24.91 & 70.42 & 35.94 & 63.32 &19.21 &64.97 &27.91 &57.13 & 42.87 & 64.92 & 26.87 & 47.47 & 35.87 & 62.42 & 29.31  & 61.38 \\
\rowcolor{myblue}
\textbf{CoCo {\footnotesize\textit{Yi-1.5B}}} & \textbf{29.93} & \textbf{73.44} & \textbf{41.12} & \textbf{67.87} &\textbf{22.78} &\textbf{66.53} &\textbf{30.04} &\textbf{59.61} & \textbf{47.74} & \textbf{67.72} & \textbf{29.03} & \textbf{49.15} & \textbf{38.92} & \textbf{64.54} & \textbf{32.04} & \textbf{64.26} \\

\midrule

RawRAG {\footnotesize\textit{DSC-1B}} & 16.36 & 65.73 & 24.58 & 53.93  &15.66 &58.68 &22.58 &50.13 & 37.31 & 59.43 & 23.75 & 43.29 & 32.00 & 58.30  & 26.25  & 57.43  \\
RepoCoder {\footnotesize\textit{DSC-1B}} & 19.55 & 67.74 & 28.29  & 56.97  &18.23 &60.15 &25.48 &52.21 & 39.44 & 61.55 & 25.38  & 45.07  & 34.31 & 60.13 & 28.63 & 59.41  \\
RLCoder {\footnotesize\textit{DSC-1B}} & 22.55 & 69.47  & 32.61  & 60.71  &20.24 &62.09 &28.61 &54.73 & 40.69 & 62.99 & 25.25  & 45.65  & 34.13 & 60.91 & 28.50  & 59.64  \\
\rowcolor{myblue}
\textbf{CoCo {\footnotesize\textbf{\textit{DSC-1B}}}}  & \textbf{26.53} & \textbf{71.97} & \textbf{36.92} & \textbf{64.87} &\textbf{23.52} &\textbf{65.37} &\textbf{32.30} &\textbf{58.62} & \textbf{44.27} & \textbf{65.91} & \textbf{27.28} & \textbf{47.36} & \textbf{37.73} & \textbf{63.84} & \textbf{30.72} & \textbf{62.69} \\

\bottomrule
\end{tabular}
}
\end{table*}
\section{Evaluation}

\subsection{RQ1: Effectiveness of Coco}
Table~\ref{tab1} presents a comparison of the overall performance of CoCo and baselines across different LLMs and benchmarks in the code completion task.
It is evident that CoCo consistently outperforms all baselines on EM, ES, ID.EM, and F1 across various backbone LLMs and benchmarks, highlighting its strong cross-model adaptability and universal effectiveness.
For example, with Yi-Coder-1.5B, CoCo improves the four metrics by 20.2\%, 4.3\%, 14,4\%, and 7.2\%  on CrossCodeEval (Python), and achieves similar gains on CrossCodeEval (Java), RepoEval (Line) and RepoEval (API), demonstrating strong robustness across diverse code distributions.
These results indicate that, beyond retrieving similar examples, incorporating  multi-granularity contextual information enables the LLM to better understand the code and generate more accurate completions.

\begin{center}
\begin{tcolorbox}[colback=gray!10,
                  colframe=black,
                  width=\linewidth,
                  arc=1mm, auto outer arc,
                  boxrule=0.5pt,
                 ]
{\textbf{Answer to RQ1:} 
CoCo consistently outperforms state-of-the-art methods across all backbone LLMs.
On CrossCodeEval (Python), it achieves improvements of 20.2\% in EM, 4.3\% in ES, 14.4\% in ID.EM, and 7.2\% in F1, demonstrating substantial gains across all key metrics.
} 
\end{tcolorbox}
\end{center}

\begin{figure}[h]
    \centering
    \includegraphics[width=0.9\linewidth]{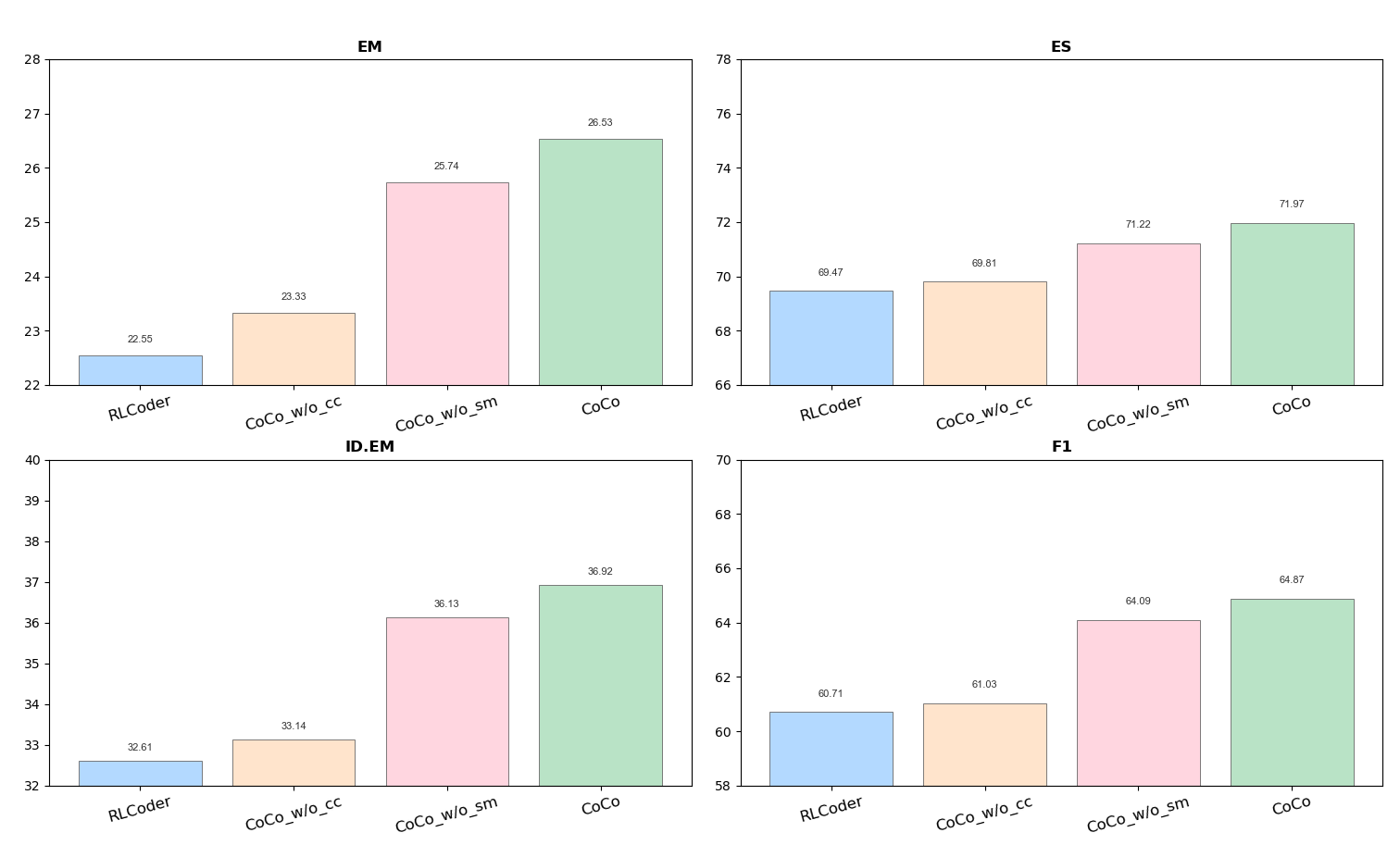}
    \caption{Performance comparison of ablation study.}
    \label{fig:ablation}
\end{figure}

\subsection{RQ2: Impact of each component.}
To assess the individual contributions of each component within CoCo framework, we conduct two ablation studies. 
We first consider \texttt{CoCo\_w/o\_cc}, a variant that removes the Code Comprehension mechanism and relies solely on semantically retrieved examples for generation. 
The second variant, \texttt{CoCo\_w/o\_sm}, disables the Structural-Aware Code Re-Ranker, so no structure-level re-ranking is performed on the retrieved candidates. 
All experiments are conducted on the CrossCodeEval (Python) benchmark using DeepSeekCoder-1B as the backbone model.

As shown in Fig.~\ref{fig:ablation}, CoCo\_w/o\_cc consistently outperforms RLCoder across all metrics, and CoCo further surpasses CoCo\_w/o\_sm, demonstrating the effectiveness of the Structural-Aware Code Re-Ranker.
While the performance gain from re-ranking is relatively modest, it still contributes positively to the final results.
This suggests that filtering semantically retrieved code examples based on structural similarity remains beneficial.
In our current design, we employ Jaccard similarity to quantify structure alignment, which may be overly restrictive. 
Future work could explore more flexible or semantically-aware structural similarity measures to improve effectiveness.
Moreover, the strong performance of CoCo\_w/o\_sm over RLCoder further confirms the importance of the Code Comprehension mechanism. 
The finding suggests that incorporating multi-granular context to enhance the model's understanding of the target code plays a critical role in improving repository-level code generation.

\begin{center}
\begin{tcolorbox}[colback=gray!10,
                  colframe=black,
                  width=\linewidth,
                  arc=1mm, auto outer arc,
                  boxrule=0.5pt,
                 ]
{\textbf{Answer to RQ2:} 
Each component contributes positively to repository-level code completion, with code comprehension having the greater impact.
} 
\end{tcolorbox}
\end{center}

\subsection{RQ3: Generalizability of CoCo}
While many existing repository-level code completion methods focus on enhancing the retriever, our proposed method, CoCo, takes a different perspective: instead of modifying the retriever, it introduces structured, multi-granularity contextual information to equip the large language model with a deeper comprehension of the unfinished code before generation, thereby improving the overall accuracy and consistency of the output.
This design enables CoCo to function as a plug-and-play enhancement module that can be seamlessly integrated into existing methods without interfering with their retrieval pipelines.
To validate the generalizability of our method, we incorporate CoCo into three representative methods—RawRAG, RepoCoder, and RLCoder—using DeepSeekCoder-1B as the backbone, and evaluate their performance on two standard benchmarks: CrossCodeEval and RepoEval.

\begin{table*}[h]
\centering
\definecolor{myblue}{rgb}{0.9, 0.95, 1.0}
\caption{Experimental results for various baselines when combined with CoCo.}
\label{tab2}
\renewcommand{\arraystretch}{1.3}
\setlength{\tabcolsep}{2pt}
\scalebox{0.9}{
\begin{tabular}{lcccccccccccccccc}
\toprule
\multirow{2}{*}{\textbf{Method}}
 & \multicolumn{4}{c}{\textbf{CrossCodeEval (Python)}} & \multicolumn{4}{c}{\textbf{CrossCodeEval (Java)}} & \multicolumn{4}{c}{\textbf{RepoEval (Line)}} & \multicolumn{4}{c}{\textbf{RepoEval (API)}} \\
\cmidrule(lr){2-5} \cmidrule(lr){6-9} \cmidrule(lr){10-13} \cmidrule(lr){14-17}
& \textbf{EM} & \textbf{ES} & \textbf{ID.EM} & \textbf{F1} & \textbf{EM} & \textbf{ES} & \textbf{ID.EM} & \textbf{F1} & \textbf{EM} & \textbf{ES} & \textbf{ID.EM} & \textbf{F1} & \textbf{EM} & \textbf{ES} & \textbf{ID.EM} & \textbf{F1} \\
\midrule

\textbf{RawRAG} & 16.36 & 65.73 & 24.58 & 53.93 & 15.66 & 58.68 & 22.58 & 50.13 & 37.31 & 59.43 & 23.75 & 43.29 & 32.00 & 58.30 & 26.25 & 57.43 \\

\rowcolor{myblue}
\textbf{RawRAG{\footnotesize\textbf{\textit{+CoCo}}}} & 22.94 & 69.08 & 32.61 & 61.47 &19.74  &60.93  &28.22  & 53.99 & 44.16 & 65.14 & 27.23 & 47.46 & 36.44 & 62.96 & 30.48 & 62.59 \\

\textbf{RepoCoder} & 19.55 & 67.74 & 28.29 & 56.97 & 18.23 & 60.15 & 25.48 & 52.51 & 39.44 & 61.55 & 25.38 & 45.07 & 34.31 & 60.13 & 28.63 & 59.41 \\

\rowcolor{myblue}
\textbf{RepoCoder{\footnotesize\textbf{\textit{+CoCo}}}} & 23.86 & 70.37 & 33.95 & 62.21 &21.51  &63.12  &30.13  & 56.38 & 43.47 & 64.85 & 26.11 & 46.82 & 36.33 & 62.99 & 29.92 & 61.88 \\

\textbf{RLCoder} & 22.55 & 69.47 & 32.61 & 60.71 & 20.24 & 62.09 & 28.61 & 54.73 & 40.69 & 62.99 & 25.25 & 45.65 & 34.13 & 60.91 & 28.50 & 59.64 \\

\rowcolor{myblue}
\textbf{CoCo} & 26.53 & 71.97 & 36.92 & 64.87 &23.52 &65.37 &32.30 &58.62 & 44.27 & 65.91 & 27.28 & 47.36 & 37.73 & 63.84 & 30.72 & 62.69 \\

\bottomrule
\end{tabular}}
\end{table*}

As shown in Table~\ref{tab2}, integrating CoCo into different baseline methods consistently results in significant performance gains across all benchmarks and evaluation metrics.
This demonstrates the general applicability of CoCo and its ability to enhance code generation quality regardless of the underlying generation backbone.
Although RepoCoder and RLCoder initially outperform RawRAG, the performance gap narrows substantially once CoCo is integrated.
Notably, in some cases, RawRAG{\footnotesize\textbf{\textit{+CoCo}}} not only closes the performance gap but even surpasses other methods, achieving the best overall results.
For example, on RepoEval~(Line) benchmark, RawRAG{\footnotesize\textbf{\textit{+CoCo}}} attains the highest F1 score among all evaluated methods.
These results suggest that multi-granularity contextual information contributes more substantially to generation quality than retrieving semantically similar examples alone, reinforcing the necessity of thoroughly modeling the underlying intent and contextual cues of the target code in repository-level generation tasks.

\begin{center}
\begin{tcolorbox}[colback=gray!10,
                  colframe=black,
                  width=\linewidth,
                  arc=1mm, auto outer arc,
                  boxrule=0.5pt,
                 ]
{\textbf{Answer to RQ3:} 
CoCo demonstrates strong generalizability when integrated into existing code generation methods, consistently enhancing performance across diverse settings. This highlights the importance of enriching the model’s comprehension of the target code, rather than relying solely on retrieved examples for guidance.
} 
\end{tcolorbox}
\end{center}

\subsection{RQ4: Time Efficiency of CoCo}

Although the previous experimental results have demonstrated the significant effectiveness of CoCo in improving code generation performance, its integration as a plug-in module inevitably introduces some computational overhead. 
To evaluate its impact on inference efficiency, we conducted comparative experiments based on RLCoder on the CrossCodeEval and RepoEval benchmarks.

As shown in Table~\ref{cost}, after integrating CoCo, the inference time increased by 132.4, 156.2, 46.0, and 72.9 on CrossCodeEval (Python), CrossCodeEval (Java), RepoEval (Line), and RepoEval (API), respectively, corresponding to 1.9\%, 2.1\%, 1.0\%, and 1.6\% increases compared to the original method.
In all cases, the overhead remains below 5\%, indicating that the integration of CoCo introduces negligible latency.
Given the consistent performance improvements it brings across benchmarks, this small cost is well justified.
\begin{table}[h]
\caption{\label{cost} Experimental Results of Time Efficiency.}
\begin{tabular}{lcc}
\hline
Benchmark             & Base Time(s) &  Overhead(s)  \\ \hline
CrossCodeEval (Python) &   7122.9           &   132.4      \\ 
CrossCodeEval (Java) &   7288.2          &   156.2     \\
RepoEval (Line)        &   4723.6         &    46.0      \\ 
RepoEval (API)         &   4667.1      &     72.9     \\ \hline
\end{tabular}
\end{table}

\begin{center}
\begin{tcolorbox}[colback=gray!10,
                  colframe=black,
                  width=\linewidth,
                  arc=1mm, auto outer arc,
                  boxrule=0.5pt,
                 ]
{\textbf{Answer to RQ4:} 
Integrating CoCo leads to a slight increase in inference latency; however, the performance gains it brings across various benchmarks justify this overhead, making the trade-off acceptable.
} 
\end{tcolorbox}
\end{center}

\subsection{RQ5: Impact of Contextual Granularity on CoCo}
To investigate how context granularity affects CoCo’s code generation performance, we conduct four ablation studies. 
The CoCo\_func variant supplies only function-level information, enriching the local execution logic of the target function. 
The CoCo\_file variant provides only file-level information to expose intra-file symbol relationships.
The CoCo\_pro variant supplies only project-level information, capturing cross-file dependencies.
Finally, CoCo\_all combines all three granularities to examine the effect of multi-granularity input.
All experiments are performed on the CrossCodeEval (Python) benchmark using DeepSeek-Coder-1B as the backbone model.
For these ablation variants, no additional context filtering is applied, allowing us to assess the impact of each granularity in its raw form.

\begin{figure}[h]
    \centering
    \includegraphics[width=0.8\linewidth]{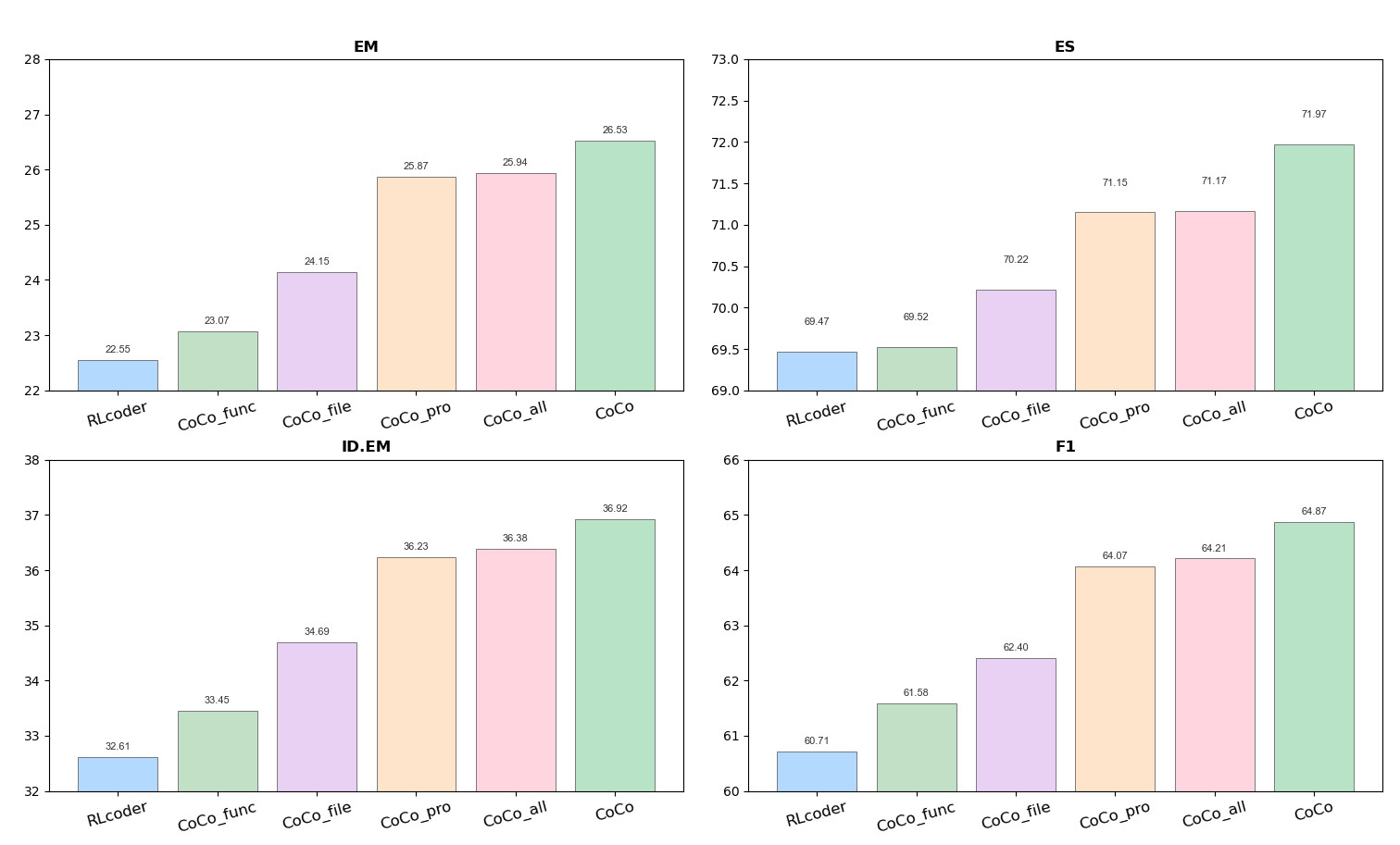}
    \caption{Performance comparison of different contextual granularity.}
    \label{fig:rq5}
\end{figure}

As shown in Fig.~\ref{fig:rq5}, all four variants outperform the RLCoder baseline, suggesting that each granularity contributes positively to code generation. 
However, the magnitude of improvement varies significantly. 
Project-level information produces the largest gains, followed by file-level information, while function-level information yields the smallest and only marginal improvement. 
This pattern is expected. 
Since function-level information mainly describes the internal logic of the target function, and the LLM already has access to the function as part of its input, it provides only incremental semantic reinforcement.
In contrast, CrossCodeEval is specifically designed for repository-level code generation, where correct output heavily depends on cross-file references.
Thus, project-level information provides the most influential external semantic cues, naturally resulting in the highest performance boost.

It is worth noting that CoCo\_all does not significantly outperform CoCo\_pro.
This indicates that simply stacking multi-granularity context does not necessarily yield substantial gains. 
While each level of contextual information contributes positively to code generation, combining all levels simultaneously does not further improve performance and may even introduce redundancy or noise, making it difficult for the model to focus on critical information in a long context. 
The superior performance of CoCo compared to CoCo\_all supports this observation. 
By employing a graph-based multi-granularity context selection mechanism, CoCo effectively filters out noise and highlights key dependencies, enabling the LLM to fully leverage information across different granularities and achieve significantly better results than any of the ablation variants. 
This further validates the effectiveness of the proposed Graph-based Multi-granularity Context Selection.

\begin{center}
\begin{tcolorbox}[colback=gray!10,
                  colframe=black,
                  width=\linewidth,
                  arc=1mm, auto outer arc,
                  boxrule=0.5pt,
                 ]
{\textbf{Answer to RQ5:} 
All granularities of contextual information positively contribute to CoCo’s code generation performance, with project-level information exerting the strongest influence. And the Graph-based Multi-granularity Context Selection effectively identifies and prioritizes key information, enabling LLMs to fully leverage multi-granularity context.
} 
\end{tcolorbox}
\end{center}

\begin{figure*}[ht]
    \centering
    \includegraphics[width=1\linewidth]{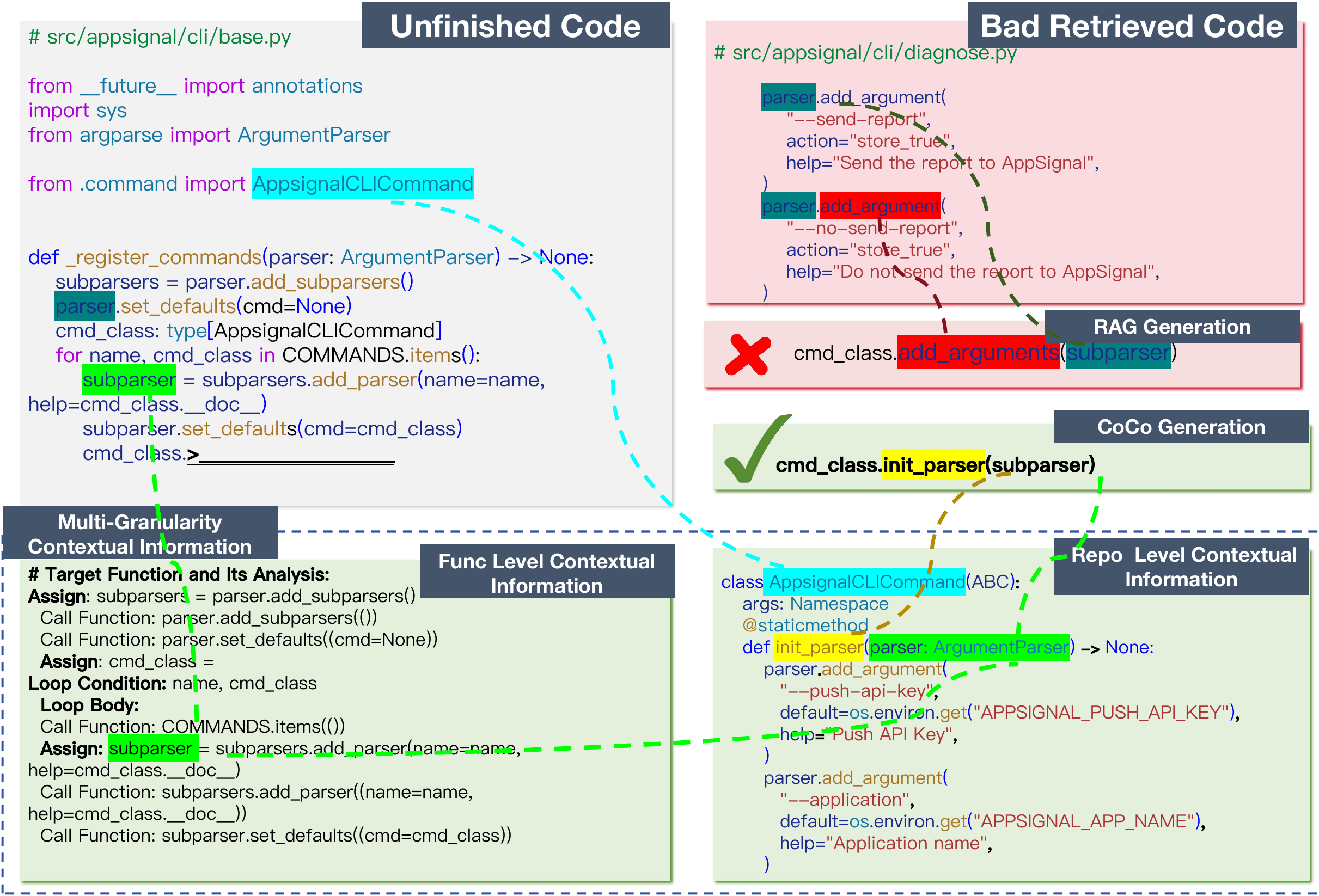}
    \caption{Case Study: Comparing CoCo with Baseline Methods}
    \label{fig:case}
\end{figure*}
\subsection{Case Study}
As illustrated in Fig.~\ref{fig:case}, this case study demonstrates the effectiveness of CoCo in enhancing code generation by leveraging multi-granular contextual information. Given a piece of unfinished code, CoCo first performs static code analysis to extract hierarchical contextual information at the function, file, and project levels.
In this example, only function-level and project-level information are informative, so file-level data is omitted for clarity.
From the function-level context, CoCo captures the local execution logic—specifically, the definition and usage of the variable \texttt{subparser} and \texttt{cmd\_class}.
The project-level context reveals that the imported class \texttt{AppsignalCLICommand} contains a mfunction named \texttt{init\_parser}, whose input parameter types are related to the variable \texttt{subparser}.
Moreover, the next line of the unfinished code references \texttt{cmd\_class} of type \texttt{AppsignalCLICommand}, suggesting that a function of this class is likely to be invoked.
Combining these cues, CoCo successfully generation code.
In contrast, RAG-based methods rely solely on surface-level semantic similarity.
In this case, the frequent occurrence of the word \texttt{parser} in both the unfinished code and candidate examples leads to the retrieval of a misleading example containing multiple \texttt{add\_argument()} calls, which subsequently biases the model toward generating incorrect code.
While CoCo also retrieves the example, its generation is further guided by the extracted multi-granular context, enabling it to ignore the misleading patterns and produce the correct completion.
This example highlights the advantage of parsering different levels of code granularity.
Unlike surface-level similarity alone, such enriched contextual comprehension allows CoCo to better infer intent and generate accurate code completions.

\section{Threats to Validity}
One potential limitation of our approach lies in the acquisition of multi-granular contextual information.
While such information enables LLMs to better understand the target code and thereby improves generation quality, it is extracted via static analysis tools using rule-based heuristics.
However, these handcrafted rules inevitably overlook certain edge cases or complex structures in practice.
For instance, the control flow graph construction may miss important execution paths in the presence of intricate logic, which can in turn affect the quality of downstream code generation. 
In future work, exploring more robust and adaptive strategies for extracting code-related context could further enhance the reliability of context construction.

Another potential limitation lies in the size and relevance of the extracted multi-granular context. 
In particular, when the parsed information includes many entire function bodies, it may significantly inflate the input length, leading to increased computational overhead and, in some cases, exceeding the model's context window. 
Moreover, not all parsed information is equally valuable.
Redundant or noisy content may distract the model and negatively impact generation performance.
To mitigate this, future research could investigate LLM-based filtering mechanisms that preselect salient contextual elements or summarize raw context to reduce noise and token overhead.
A further limitation of our work is its primary focus on the generation of substantive code entities—such as functions, variables, and control structures—while largely overlooking non-entity code elements that are commonplace in real-world software engineering. In practical repository-level code completion scenarios, developers frequently need to generate or complete non-entity code components, including comments, annotations, and formatting symbols~\cite{fakhoury2019improving,tenny1988program}. These elements, while not directly contributing to program logic, are essential for code readability and maintainability. Future work could extend our method to better support and assess the generation of these elements.

\section{Related Work}
\subsection{Repository Level Code Completion}
Repository-level code completion is highly significant as it better reflects the complexities and requirements of real-world software development compared to traditional, function-level code completion. This technique is inherently flexible and can be seamlessly integrated into modern IDEs and programming plugins~\cite{copilot2023github}, making advanced code intelligence readily accessible in daily development workflows. 
Early efforts to enhance neural code generation models focused on incorporating code structure information, such as leveraging ASTs and static analysis tools. For example, AST-T5~\cite{gong2024ast} utilizes a T5 model enhanced by AST-guided segmentation and span corruption objectives to integrate code structural information during pretraining. Similarly,~\cite{jiang2021ast} proposes enhancing a tree-structured LSTM decoder by introducing AST-based attention over historical actions and applying multi-task learning for joint prediction of current and future actions.
While an increasing number of large language models~(LLMs) have demonstrated strong performance on general code completion tasks, they often struggle with repository-level code completion due to a lack of repository-specific knowledge~\cite{ding2023crosscodeeval,liu2023repobench,tang2023domain}.
Repository-level code completion is attracting significant attention as a key challenge for intelligent software development in real-world scenarios~\cite{ding2022cocomic,liao2023context,shrivastava2023repofusion,pei2023better}.
To address this limitation, it is crucial to effectively inject repository knowledge into LLMs.
For example, CoCoMIC~\cite{ding2022cocomic} and RepoFusion~\cite{shrivastava2023repofusion} finetune LLMs using both in-file and relevant cross-file contexts, thereby incorporating repository knowledge into the models. However, such methods are generally limited to open-source models. 
To address the problem, some post-processing frameworks based on pre-trained models~\cite{khandelwal2019generalization,tang2023domain} have been introduced. 
These frameworks adjust the output probability of the next token prediction based on the token frequency statistics in the repository.

\subsection{Retrieval-Augmented Code Completion}
With the rise of Retrieval-Augmented Generation~(RAG), it has become a common paradigm for repository-level code completion.
The core idea is to retrieve the top-k semantically similar code snippets from the repository based on the target code, and use them as examples to guide large language models (LLMs) in generation~\cite{liao2023context,lu2022reacc,tan2024prompt,zhang2023repocoder}. 
However, conventional RAG methods often treat code as plain text during retrieval, neglecting its inherent structural and semantic characteristics. 
In order to overcome the limitations of conventional RAG methods, a growing number of approaches have been suggested to improve the quality of generation by enhancing the retriever, such as incorporating structural information or syntax-aware techniques.
For example, DraCo~\cite{cheng2024dataflow} introduces a dataflow-guided retrieval strategy by parsing repositories into fine-grained code entities and constructing a repository-specific context graph, enabling more accurate and structurally aware retrieval than traditional text-based methods.
GraphCoder~\cite{liu2024graphcoder} introduces a structured, graph-based retrieval framework that leverages code context graphs to more effectively identify and weigh relevant code snippets for completion tasks, substantially improving retrieval quality and efficiency. 
RRG~\cite{gao2024preference} introduces a code refactorer module between retrieval and generation, which compresses and restructures retrieved code into more model-friendly and answer-relevant contexts, thereby improving code generation quality and reducing inference costs.
Repoformer~\cite{wu2024repoformer} proposes a selective retrieval strategy, allowing the model to decide whether or not to retrieve external context for a given input. This conditional retrieval mechanism reduces unnecessary overhead and improves robustness.
RLCoder ~\cite{wang2024rlcoder} introduces a reinforcement learning-based retriever that autonomously learns to select and filter useful context for code generation without supervision, leveraging feedback from code perplexity and a stop signal mechanism, thereby achieving significant performance gains over state-of-the-art baselines.

\section{CONCLUSION}
In this paper, we propose CoCo, a novel framework for repository-level code completion.
Unlike other methods that primarily focus on improving RAG-based retrieval quality, CoCo takes a complementary perspective by enhancing the model’s comprehension of unfinished code before generation. 
Specifically, it leverages static analysis tools to extract multi-granularity contextual information—including function-level, file-level, and project-level semantics—capturing both the potential execution logic and relevant dependencies.
Experimental results demonstrate that CoCo consistently outperforms state-of-the-art baselines across multiple benchmarks and exhibits strong generalizability, enabling seamless integration into other code generation methods to further enhance performance.
\section{Acknowledgments}



\bibliographystyle{ACM-Reference-Format}
\bibliography{sample-base}

\end{document}